# The sub-critical illusion: synthetic Zeeman effect observations from galactic zoom-in simulations

Zipeng Hu ,[1]★ Benjamin D. Wibking [1,2] and Mark R. Krumholz [1,2]

[1]*Research School of Astronomy and Astrophysics, Australian National University, Canberra, ACT 2611, Australia*
[2]*ARC Centre of Excellence for Astronomy in Three Dimensions (ASTRO-3D), Canberra, ACT 2611, Australia*

## ABSTRACT

Mass-to-flux ratios measured via the Zeeman effect suggest the existence of a transition from a magnetically sub-critical state in H I clouds to a supercritical state in molecular clouds. However, due to projection, chemical, and excitation effects, Zeeman measurements are subject to a number of biases, and may not reflect the true relations between gravitational and magnetic energies. In this paper, we carry out simulations of the formation of magnetized molecular clouds, zooming in from an entire galaxy to sub-pc scales, which we post-process to produce synthetic H I and OH Zeeman measurements. The mass-to-flux ratios we recover from the simulated observations show a transition in magnetic criticality that closely matches observations, but we find that the gravitational-magnetic energy ratios on corresponding scales are mostly supercritical, even in the H I regime. We conclude that H I clouds in the process of assembling to form molecular clouds are already supercritical even before $H_2$ forms, and that the apparent transition from sub- to supercriticality between H I and $H_2$ is primarily an illusion created by chemical and excitation biases affecting the Zeeman measurements.

## 1 INTRODUCTION

The dynamical importance of magnetic fields (B-field) relative to gravity during the star formation process remains an open question. Quantitatively, for a uniform spherically symmetric cloud of mass $M$, volume $V$, and radius $R$, threaded by a uniform magnetic field of intensity $B$, the ratio of magnetic to gravitational energy is

$$\frac{E_{\text{mag}}}{E_{\text{grav}}} = \frac{B^2 V}{8\pi} \times \frac{3R}{5GM^2} \propto \frac{B^2 R^4}{5GM^2} \propto \left(\frac{\Phi}{M}\right)^2, \quad (1)$$

where $\Phi = \pi R^2 B$ is the magnetic flux, which is conserved under ideal magnetohydrodynamics (MHD). Mouschovias & Spitzer (1976) show that clouds can collapse only if their mass-to-flux ratio $M/\Phi$ exceeds a critical value

$$\left(\frac{M}{\Phi}\right)_{\text{crit}} = \frac{1}{3\pi}\sqrt{\frac{5}{G}}. \quad (2)$$

A cloud with $\mu_B \equiv (M/\Phi)/(M/\Phi)_{\text{crit}} > 1$ is called supercritical, implying a weak magnetic field compared to gravity, while a cloud with $\mu_B < 1$ has strong magnetic field support against gravity and is called sub-critical. In a sub-critical cloud, the magnetic field strength will rapidly grow during a uniform gravitational collapse, preventing further contraction long before stars form (Mouschovias & Spitzer 1976). Thus if the bulk of the interstellar medium (ISM) is magnetically sub-critical, non-ideal MHD effects are required to allow stars to form. One candidate effect is the ambipolar diffusion (Shu 1977; Shu, Adams & Lizano 1987), whereby in the very poorly ionized regions found at high densities, neutral gas becomes decoupled from ions, allowing it to cross the field lines and further collapse. Alternately, turbulent motion of the gas might lead to magnetic reconnection and change the $B$-field topology, releasing magnetic flux and enabling collapse (Vishniac & Lazarian 1999). Understanding whether such non-ideal processes are needed, and if so which ones are dominant, requires observational work to precisely measure the magnetic field structure both in and around molecular clouds.

★ E-mail: zphu.charles@gmail.com

The most direct method of measuring magnetic fields, and by far the least uncertain when it comes to measuring field strengths, is via the Zeeman effect, which reveals the strength of magnetic fields along the line of sight (LOS) ($B_{\text{LOS}}$) via the frequency splitting these fields induce between the components of emission or absorption lines with different polarizations. Using a range of Zeeman-sensitive tracers (most commonly H I, OH, and CN), the method can cover a large range of total hydrogen column density ($N_{\text{H}}$). Crutcher (2012) summarizes the results of five Zeeman effect surveys (Crutcher et al. 1999; Bourke et al. 2001; Heiles & Troland 2004; Falgarone et al. 2008; Troland & Crutcher 2008), which show a clear transition at $N_{\text{H}} = 10^{21-22} \text{ cm}^{-2}$: diffuse H I regions with lower column densities are mostly sub-critical, while OH and CN measurements that trace denser regions indicate that they are mostly supercritical. Troland et al. (2016) make much more precise Zeeman effect measurements with both H I and OH tracers in the Orion Veil, and Thompson, Troland & Heiles (2019) conduct an OH absorption line Zeeman effect survey in the molecular cloud envelopes. Both these studies find results consistent with the Crutcher (2012) review, showing a sub-critical to supercritical transition in the same column density range. The mechanism behind such a transition, however, is still unclear, partially due to the lack of observations in the atomic-molecular transition regime. In an early attempt to explore this regime, Ching et al. (2022) perform H I narrow self-absorption (HINSA) Zeeman effect observation on L1544, and their result suggests that the cloud is already supercritical when $H_2$ molecules start to form.

Partly in an attempt to probe the sub- to supercritical transition, quite a few theoretical works have also investigated magnetic fields in the cloud formation process using MHD simulations. Bertram et al. (2012) determine the mass-to-flux ratios in cores (∼0.1 pc) and clumps (∼1 pc) from their simulations starting with a range of initial magnetic field strengths. They find that the mass-to-flux ratio of the structures they form is only loosely related to the initial criticality of the large-scale environment. The recent work by Ibáñez-Mejía, Mac Low & Klessen (2022) starts from a galactic-scale simulation with a box size of 1 kpc, then zooms into three dense clouds, reaching a maximum resolution of 0.1 pc. They report the criticality transition happens when the hydrogen number density $n_H \approx 100$ cm$^{-3}$. Although these simulations produce results similar to observations, it is still unclear how the sub- to supercritical transition occurs, particularly since their simulations use ideal MHD. Priestley, Yin & Wurster (2022) produce synthetic molecular line observations from the pre-stellar cores in their non-ideal MHD simulations, and argue that magnetically sub-critical cores can actually appear supercritical in observations.

One significant limit of these works, however, is that they begin from an arbitrarily prescribed uniform magnetic field morphology, rather than a realistic galactic magnetic field structure. Given that observations indicate that magnetic fields are continuously connected from galactic to cloud scales (e.g. Li et al. 2014), this is a potentially significant issue. Moreover, none of the works published to date have solved the radiative transfer problem when producing H I 21 cm line synthetic observations, or fully solved the problem of non-local thermodynamic equilibrium (non-LTE) excitation for the Zeeman-sensitive OH lines. Since the regions of interest are those where H I self-absorption is expected to be strongest (as illustrated by the Ching et al. 2022 HINSA observations), and where the density is well below the OH critical density, this is another potentially significant limitation.

In this paper, we study the transition between the sub-critical and supercritical ISM using ideal MHD simulations that start from the scale of the whole galaxy and then zoom in to reach maximum resolutions in dense regions up to ≈0.4 pc. These simulations allow us to for the first time combine realistic galactic magnetic field structures with detailed dynamics on the cloud scale in a single self-consistent calculation. We post-process our simulations to derive realistic chemical compositions, and we then solve the radiative transfer equation including non-LTE excitation effects along different LOSs to produce synthetic H I and OH observations. Our goal is to understand the extent to which observed mass-to-flux ratio accurately reflect the ratio of magnetic to gravitational energy in the 3D volume they probe, and, to the extent that they do so, to understand the sub- to supercritical transition that appears in the observations arises in the true 3D structure of the gas and magnetic field as they assemble to form molecular clouds.

This paper is structured as follows: Section 2 summarizes the numerical method and simulation data. Section 3 describes our post-processing analysis method to derive chemical composition, calculate radiative transfer, and produce the synthetic observations. Section 4 presents the comparison between synthetic observations and true 3D energy relations. Section 5 discusses the physical meaning behind the results, and possible future work in this area. Section 6 concludes the work done in the paper.

## 2 SIMULATION

Our work begins from a simulation of an entire galaxy. To produce synthetic OH Zeeman effect observations from this, we need to locate molecular clouds with strong OH emission. Then we can produce synthetic H I observations in the atomic regions surrounding the selected molecular clouds. To achieve this goal, the scheme of the simulations in this work has three main steps: (1) carry out MHD simulations of a full galaxy; (2) identify regions with strong OH emission; (3) re-simulate the formation history of the selected regions with higher resolution. We discuss both the numerical method and the detailed scheme of simulations in this section.

### 2.1 Numerical method

Our full-galaxy simulations are based on the galactic magnetic field structure simulation by Wibking & Krumholz (2022, hereafter WK22). Here, we simply summarize the physics and numerical methods, and highlight areas where our treatment differs from that of WK22. For details on the remainder of the methods, we refer readers to WK22. We solve the equations of ideal MHD with the GIZMO code (Hopkins 2015, 2016; Hopkins & Raives 2016), which uses a mesh-free, Lagrangian finite-mass Godunov method. The simulations include a time-dependent chemistry network that tracks the abundances of H I, H II, He I, He II, He III, and free electrons, and uses these abundances to calculate the atomic cooling rates due to hydrogen and helium. We also include metal line cooling, using cooling rates interpolated from a prescribed table as a function of density and temperature under the assumption of ionization equilibrium and solar metallicity. The table is computed with CLOUDY (Ferland et al. 1998), using the method of Smith, Sigurdsson & Abel (2008) as applied in the GRACKLE chemistry and cooling library (Smith et al. 2017).

For gas particles with density $\rho_g$ larger that a threshold $\rho_{crit}$, we calculate a local star formation rate density

$$\rho_{SFR} = \epsilon_{ff} \frac{\rho_g}{t_{ff}}, \tag{3}$$

where $\epsilon_{ff}$ is the star formation efficiency, $\rho_g$ is the local gas density, and $t_{ff}$ is the local gas free-fall time. As summarized in the review by Krumholz, McKee & Bland-Hawthorn (2019), $\epsilon_{ff}$ has a mean value ≈0.01 across a wide density range, which is also the value we adopt. We set the value of $\rho_{crit}$ to be approximately equal to the Jeans density at our mass resolution; in the simulations of WK22, the mass resolution $\Delta M = 859.3$ M$_\odot$, and we therefore set $\rho_{crit} = 100$ m$_H$ cm$^{-3}$, but we increase the mass resolution and thus $\rho_{crit}$ in the zoom-in stages as described below. If $\rho_g > 100\rho_{crit}$, we increase $\epsilon_{ff}$ to 1 in order to limit the computational cost, since high densities require correspondingly small time-steps. This effectively sets an upper limit of $100\rho_{crit}$ on the gas density resolved by the simulation, since gas particles with higher density will instantly be converted into stellar particles. Gas particles for which $\rho_{SFR} > 0$ have probability

$$P = 1 - \exp(-\epsilon_{ff}\Delta t/t_{ff}) \tag{4}$$

of being converted to a star particle per time-step of size $\Delta t$. Stellar particles represent clusters that only interact with the galactic environment via gravity and feedback.

While our treatment of star formation and cooling is identical to that of WK22 (other than our use of a higher $\rho_{crit}$ at higher resolution), our treatment of stellar feedback is necessarily slightly different because our resolution will be high enough that stellar particles no longer represent clusters massive enough that we expect them to sample the full initial mass function (IMF). This matters because it means that we cannot compute stellar feedback simply by integrating over the IMF. Instead, for each stellar particle, we stochastically draw its stellar population from a Chabrier IMF (Chabrier 2005) through the stellar population synthesis code SLUG (da Silva, Fumagalli & Krumholz 2012; Krumholz et al. 2015), following the approach to using SLUG as a sub-grid model described by Fujimoto, Krumholz & Tachibana (2018) and Armillotta et al. (2019). We compute stellar evolution using the Padova stellar tracks (Bressan et al. 2012), and derive stars' ionizing luminosities from their masses and ages using the 'STARBURST99' spectral synthesis method (Leitherer et al. 1999). We determine which stars end their lives as Type II supernovae, and the stellar ages at which those supernovae occur, using the tabulated results of Sukhbold et al. (2016). Our numerical treatment of ionization and supernova feedback is identical to that of WK22; our method in this paper differs only in how we determine the ionizing luminosity, number, and timing of supernovae produced by a given stellar particle.

## 2.2 Zoom-in method

The original WK22 simulation follows a Milky Way-like galaxy at a mass resolution $\Delta M = 859.3$ M$_\odot$, insufficient to resolve the small, dense regions from which the OH lines used in Zeeman effect measurements arise. In order to capture these regions, we carry out two additional simulation stages to zoom into molecular clouds from the galactic scale. For the first stage, we use the $t = 0.6$ Gyr snapshot in WK22's MHD simulation as our initial condition. We choose this snapshot because both the mass-weighted mean magnetic field strength (6 μG) and total SFR (2 M$_\odot$ yr$^{-1}$) are stable and close to the values observed for the present-day Milky Way. WK22 explore the properties of the simulated galaxy at this time in detail, so we will not discuss them further here.

As a first step towards increasing the resolution, at this point we switch the feedback prescription over to the SLUG-based method described in Section 2.1 and continue the simulation at the same resolution for another 0.1 Gyr. During this time the SFR undergoes a transient increase, since no supernovae occur for ≈4 Myr after the change in feedback prescription, but thereafter the SFR stabilizes at around 3 M$_\odot$ yr$^{-1}$ for the last 0.05 Gyr. Once the simulation reaches 0.7 Gyr of evolution, our next zoom-in step is that we replace each gas particle with 10 smaller particles, which are randomly distributed around the centre of mass of their parent particle with a distance of 0.2 times the parental smoothing length. All other physical parameters remain the same during the split, except that the particle mass is divided by 10, and the smoothing length is divided by $10^{1/3}$. In this way we increase the mean gaseous mass resolution to ≈80 M$_\odot$. During this stage we also impose $\rho_{crit}$ to $10^4$ m$_H$/cm$^3$, which is approximately the Jeans density at the mass resolution we will reach during the next zoom-in stage; we use this value rather than the Jeans density during this simulation stage to ensure that there are no transients after we zoom in again. We continue this mid-resolution run for another 21 Myr, to $t = 0.721$ Gyr. We show the face-on gas surface density $\Sigma_{gas}$ at this time in Fig. 1.

In order to select regions for further zoom-in, we use the chemistry post-analysis code DESPOTIC to derive the OH 1665 and 1667 MHz emission line luminosities of each gas particle; we provide details of our method for doing so in Section 3.1. For each gas particle we first determine its total luminosity $L_{OH}$ as the sum of these two lines, then randomly select seven particles with $L_{OH} > (2/3)\max(L_{OH})$, where the maximum is computed over all gas particles. We then draw a sphere with a radius of 0.3 kpc around each selected particle; we show these regions as white circles in Fig. 1. We choose this radius to be close to the gaseous disc scale height 0.34 kpc. The free-fall times of the seven spheres range from 1.2 to 21 Myr, with a median value of 3.8 Myr. Therefore, simulating the last 10 Myr of these regions will be sufficient to reconstruct their formation history.

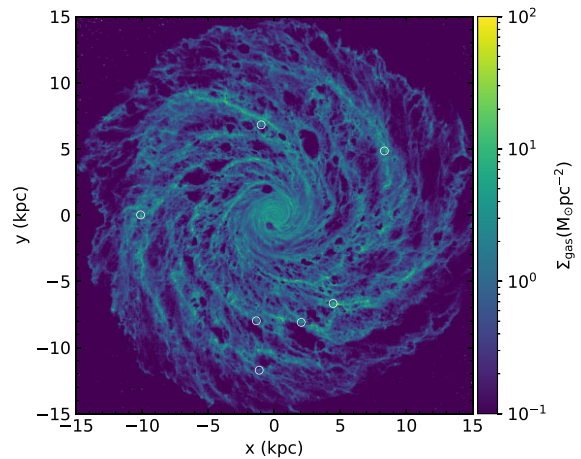

**Figure 1.** Face-on gas surface density map of the last snapshot ($t = 0.721$ Gyr) of the mid-resolution run. The seven white circles, each of radius 0.3 kpc, are the high OH luminosity regions we select for the highest resolution simulation stage.

We record the particle IDs of all gas particles within the seven spheres, then find them in an earlier simulation snapshot 10 Myr before the one shown in Fig. 1 ($t = 0.711$ Gyr). Using the method described above, we split these gas particles in 10 again, reaching a mass resolution of 8.9 M$_\odot$; the corresponding Jeans length is ≈0.4 pc, which we can use as an estimate of the spatial resolution in the dense regions. We then re-simulate the last 10 Myr ($t = 0.711$–0.721 Gyr), so that the regions we will use in our final analysis have all been simulated for ≈1 free-fall time at our highest resolution. We refer to this final stage as the high-resolution run.

To check if the simulation result is consistent after splitting selected gas particles, we compare the face-on surface density maps of the mid- and high-resolution runs at $t = 0.721$ Gyr for one example region in Fig. 2. We see that, as expected, the large-scale density structure is the same, but that the higher resolution case shows more structure at high $\Sigma_{gas}$.

To derive some basic physical parameters on the molecular cloud scale (∼100 pc), we draw (100 pc)$^3$ cubes around the densest particle for each group of split particles in the last high-resolution snapshot. This size is close to the typical size of the largest molecular cloud complexes (Krumholz 2014b). In Table 1, we tabulate the total mass in the form of H I, $M_{HI}$, total mass in H$_2$, $M_{H_2}$, volume-weighted mean hydrogen number density, $\bar{n}_H$, and volume-weighted and mass-weighted mean magnetic field strengths, $\bar{B}_V$ and $\bar{B}_M$, in each of these seven cubes at $t = 0.721$ Gyr. We explain how we derive $M_{HI}$ and $M_{H_2}$ in Section 3.1.

# 3 SYNTHETIC OBSERVATIONS

To study the mechanism behind the observed mass-to-flux ratio transition, we need to produce synthetic observations using methods that match real observational methods as closely as possible. This process consists of three main steps: (1) determine the chemical state of each gas particle, and its absorption and emission coefficients in the H I and Zeeman-sensitive OH lines, respectively; (2) solve the equation of radiative transfer to obtain synthetic spectra; (3) produce synthetic emission maps and B-field maps from the spectra, using the same techniques used to analyse observed spectra. We now proceed to describe these steps in detail.

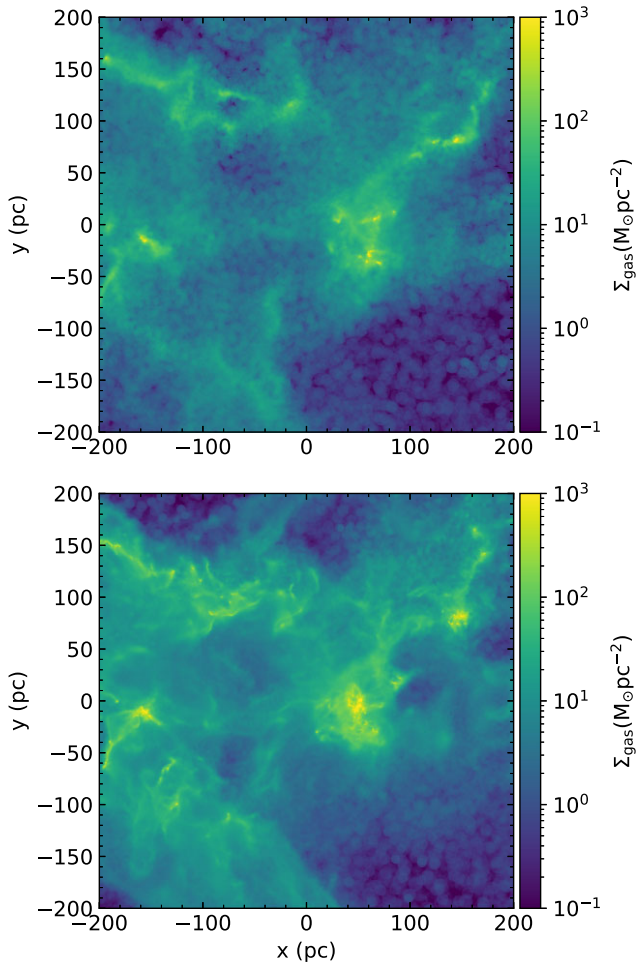

**Figure 2.** Face-on gas surface density maps of cloud 4 in the mid-resolution (top panel) and high-resolution (bottom panel) simulations at time $t = 0.721$ Gyr.

### 3.1 Chemistry and excitation calculation

Since our simulation does not include molecular chemistry, and even simulations that do include chemistry (e.g. Glover et al. 2010) do not self-consistently compute non-LTE excitation (which is critical for the OH emission), we post-process our simulations using the DESPOTIC astrochemistry and radiative transfer code (Krumholz 2014a). DESPOTIC derives the equilibrium excitation, thermal, and chemical states of clouds from their elemental abundances (assumed to be Solar for this work), radiation field environment, and physical properties. The thermal equilibrium calculation includes cosmic ray heating, molecule line cooling, the grain photoelectric effect, and collisional energy exchange between gas and dust, while the chemical equilibrium calculation adopts the H-C-O chemical network by Gong, Ostriker & Wolfire (2017). The excitation calculation, which also determines optical depths for line cooling, uses the large velocity gradient approximation. We refer the readers to Krumholz (2014a) for more details on DESPOTIC.

We perform the chemistry post-analysis following the methodology of Armillotta, Krumholz & Di Teodoro (2020). The first step is to generate tabulations of the H I and H$_2$ mass fractions, $X_{HI}$ and $X_{H_2}$, the OH transition line luminosities at 1665 and 1667 MHz, $L_{OH, 1665}$ and $L_{OH, 1667}$, and the gas temperature[1] $T$, as a function of local velocity gradient, d$v$/d$r$, hydrogen number density, $n_H$, and hydrogen column density, $N_H$; note that $n_H$ and $N_H$ are the number of hydrogen nuclei per unit volume and per unit area, so these quantities are invariant under different chemical compositions. Since all seven clouds are on spiral arms away from the galactic centre, we assume the interstellar radiation field (ISRF) strength to be $\chi = 1 \, G_0$, where $1 \, G_0$ corresponds to the solar neighbourhood ISRF strength (Draine 1978). We set the primary cosmic ray ionization rate to a Solar neighbourhood-like value $\zeta = 10^{-16} \text{ s}^{-1}$. We run DESPOTIC on a grid of models covering the parameter range $n_H = 10^{-2} - 10^6 \text{ cm}^{-3}$, $N_H = 10^{19} - 10^{25} \text{ cm}^{-2}$, and d$v$/d$r$ $= 10^{-3} - 10^2 \text{ km s}^{-1} \text{ pc}^{-1}$. The final grid is $21 \times 16 \times 15$ points in size in $(n_H, N_H, dv/dr)$ space, and is uniformly spaced in logarithm. We have verified by visual inspection that quantities of interest vary smoothly over our grid, so interpolation errors are small.

**Table 1.** Physical parameters derived from $(100 \text{ pc})^3$ cubes drawn around the centres of mass of the high-resolution regions at simulation time $t = 0.721$ Gyr. $M_{HI}$ and $M_{H_2}$ are the total mass in the form of H I and H$_2$, respectively. $\bar{n}_H$ is volume-weighted mean hydrogen number density. $\bar{B}_V$ and $\bar{B}_M$ refer to volume-weighted and mass-weighted mean magnetic field strength, respectively.

| Cloud | $M_{HI}$ ($10^5 \text{ M}_\odot$) | $M_{H_2}$ ($10^5 \text{ M}_\odot$) | $\bar{n}_H$ (cm$^{-3}$) | $\bar{B}_V$ ($\mu$G) | $\bar{B}_M$ ($\mu$G) |
|---|---|---|---|---|---|
| 1 | 0.35 | 0.23 | 5.0 | 2.7 | 5.1 |
| 2 | 0.26 | 0.54 | 5.8 | 1.4 | 4.9 |
| 3 | 0.35 | 0.72 | 6.6 | 1.2 | 9.9 |
| 4 | 0.55 | 0.75 | 8.3 | 1.2 | 7.9 |
| 5 | 1.1 | 1.4 | 16 | 15 | 22 |
| 6 | 1.0 | 3.0 | 20 | 7.2 | 20 |
| 7 | 1.6 | 10 | 53 | 17 | 48 |

The second step is to assign values of $X_{HI}$, $X_{H_2}$, $L_{OH, 1665}$, $L_{OH, 1667}$, and $T$ to each gas particle in the simulations by performing trilinear interpolation in the variables d$v$/d$r$, $n_H$, and $N_H$. We compute the hydrogen number density for each particle as

$$n_H = \frac{X_H \rho_g}{m_H},$$ (5)

where $X_H = 0.71$ is the mass fraction of hydrogen assuming Solar metallicity, $\rho_g$ is the local gas density, and $m_H = 1.67 \times 10^{-24}$ g is the mass of the hydrogen nucleus. For $N_H$, we adopt the definition in Safranek-Shrader et al. (2017) as

$$N_H = n_H L_{shield},$$ (6)

where $L_{shield}$, named as shielding length, represents the approximate length that should be used in estimating the column density of material that shields the gas against the dissociating ISRF. Safranek-Shrader et al. (2017) show that the local approximation to the shielding length that most closely reproduces the results of detailed ray-tracing radiative transfer calculations is to set $L_{shield}$ to the local

---

[1] Note that the temperature returned by the DESPOTIC calculation does not match the temperature found by GIZMO because the GIZMO calculation does not include molecular cooling, while DESPOTIC does; thus DESPOTIC generally yields lower temperatures in dense, molecular regions. This difference is unimportant for the dynamics, because in such cold regions thermal pressure is generally an unimportant force, but it matters a great deal for the excitation and radiative transfer calculations.

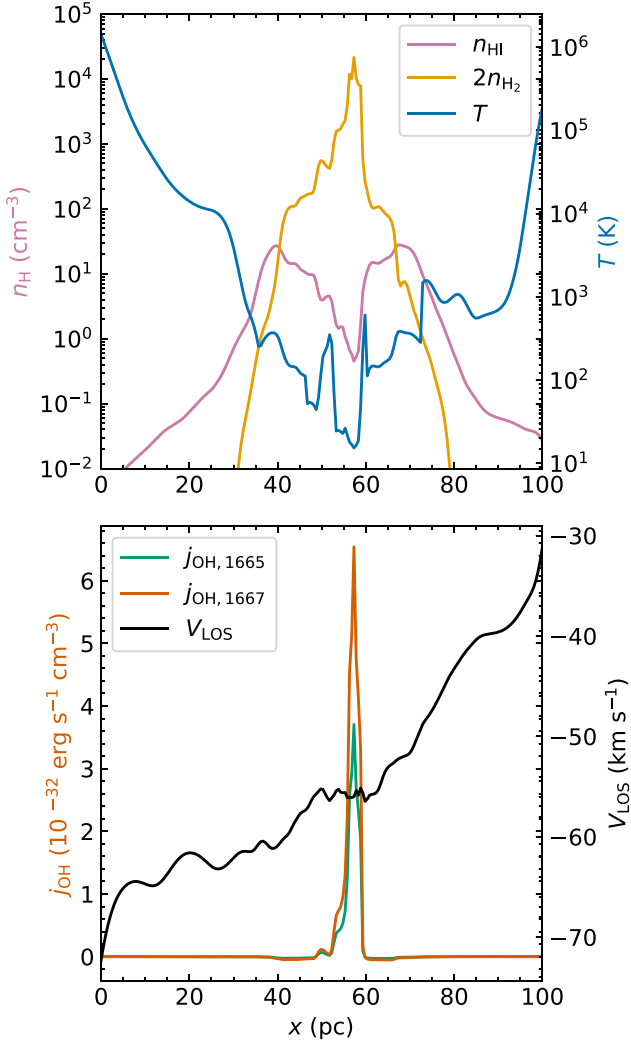

**Figure 3.** Gas properties along a 100 pc ray through the point of the lowest gas temperature for cloud 4; the ray is parallel to the $x$-axis shown in Fig. 2. Top panel: number density of H nuclei $n_H$ in the form of H I (pink) and H$_2$ (orange), together with local gas temperature $T$ (blue). Bottom panel: frequency- and angle-integrated emissivity $j_{OH, 1665}$ (green line) and $j_{OH, 1667}$ (red line) in the OH 18 cm lines, and line-of-sight velocity $V_{LOS}$ (black line).

Jeans length $L_J$, with the Jeans length computed using the smaller of the local gas temperature and $T = 40$ K. Finally, we calculate the velocity gradient as

$$\frac{dv}{dr} = \|\nabla \times v\|, \tag{7}$$

where $v$ is the local gas velocity. Our grid covers the full range of gas particle properties in the high-resolution simulations, so extrapolation off the grid is not necessary.

To illustrate the outcome of our interpolation procedure, in Fig. 3 we plot the distribution of gas properties along a 100 pc ray through the minimum of gas temperature in cloud 4. This ray is along the $x$-axis as shown in Fig. 2, and we set its coordinates in the transverse direction. In Fig. 3, we sample the ray at 200 positions spaced at intervals of $\Delta L = 0.5$ pc. We compute the properties shown at each sampling point as a sum over nearby gas particles, weighted by their smoothing kernels following the GIZMO kernel weighting scheme described by Hopkins (2015). The upper panel shows the H I number density $n_{HI}$ (pink), twice the H$_2$ number density $2n_{H_2}$ (orange), and the gas temperature $T$ (blue), while the bottom panel shows the angle-averaged, frequency-integrated emissivity in the OH 18 cm lines $j_{OH, 1665}$ (green) and $j_{OH, 1667}$ (red), together with line-of-sight velocity $V_{LOS}$ (black), where we take the LOS direction to be along the ray. In these plots there is a clear transition from H I to H$_2$ in the central cold region, and we see that OH emission is confined to this region.

### 3.2 Radiative transfer

Our next step is to compute observable spectra. For OH this is straightforward, since Zeeman measurements are made using OH lines seen in emission, and observations indicate that the OH lines used are invariably optically thin. Thus we can compute the OH intensity as a function of velocity simply by integrating the emissivities returned by the DESPOTIC calculation along the LOS. In practice, we do this by interpolating the particle data on to a $200^3$ cube with a $\Delta L = 0.5$ pc voxel size centred on the densest particle; at each voxel we compute the OH emissivities in the 1665 and 1667 MHz lines by interpolating the particle data as described in Section 3.1 and used to construct Fig. 3. We now consider LOSs through each pixel on the face of the cube in the $x$ and $y$ directions, i.e. the two directions that lie within the galactic plane; thus we have a total of $2 \times 200^2$ pixels for each of our seven clouds. For each pixel, we determine its mass-weighted mean line-of-sight velocity $\bar{V}_{LOS}$, and consider a grid of velocities over a range $\bar{V}_{LOS} \pm 10$ km s$^{-1}$ with 0.1 km s$^{-1}$ resolution. We then integrate the intensity in each pixel and at each velocity to produce our final OH spectra.

The situation for H I is more complex, because H I Zeeman measurements are generally made using absorption against background radio continuum sources (e.g. Heiles & Troland 2004), and comparing the on-source absorption profile against an off-source emission profile. We therefore require a full radiative transfer calculation for H I. Our first step in this calculation is to interpolate the particle data on to a regular cube exactly as for OH; at each pixel we have the H I and H$_2$ number densities $n_{HI}$ and $n_{H_2}$, along with the gas temperature $T$, gas velocity $v$, and the local sound speed $c_s = \sqrt{k_B T/m_H}$. Then we compute the H I emissivity $j_v$ and attenuation coefficient $\kappa_v$ as

$$j_v = \frac{3}{16\pi} A_{ul} E_{ul} n_{HI} \phi_v, \tag{8}$$

$$\kappa_v = \frac{3 A_{ul} hc \lambda_{ul} n_{HI}}{32\pi k_B T} \phi_v, \tag{9}$$

where $\lambda_{ul} = 21.12$ cm is the transition rest wavelength, $E_{ul} = 5.87 \times 10^{-6}$ eV is the transition energy, $A_{ul} = 2.88 \times 10^{-15}$ s$^{-1}$ is the Einstein coefficient, and $\phi_v$ is the line shape function, given by

$$\phi_v = \frac{\lambda_{ul}}{\sqrt{2\pi} c_s} e^{-(V_c - V_{LOS})^2/c_s^2}, \tag{10}$$

where $V_c$ is the channel velocity and $V_{LOS}$ is the component of the gas velocity $v$ parallel to the LOS.

To produce absorption spectra against background radio sources, we assume that H I emission is negligible, and add up the optical depth along each LOS and for each velocity channel as

$$\tau_v = \sum_i \kappa_{v,i} \Delta L, \tag{11}$$

where $i$ denotes the pixel number, increasing from $i = 1$ for the farthest pixel to $i = 200$ for the nearest one. To produce the corresponding H I emission spectra for an off-source observation, we solve the radiation transfer equation along the same lines of sight

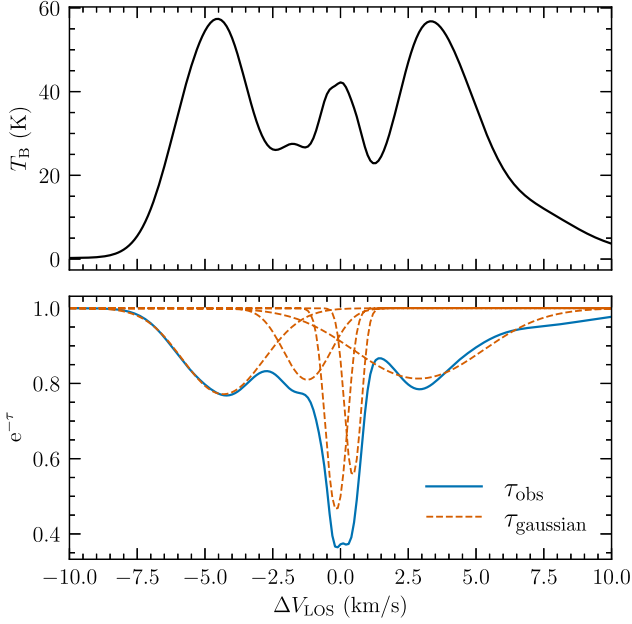

**Figure 4.** Top panel: the synthetic H I emission spectrum (off-source) corresponding to the LOS illustrated in Fig. 3. $T_B$ is the brightness temperature corresponding to the observed intensity, and $\Delta V_{LOS}$ is the difference between the velocity channel and the mass-weighted mean $\bar{V}_{LOS}$ of the whole LOS. The velocity resolution is $0.1 \text{ km s}^{-1}$. Bottom panel: absorption depth from the same LOS assuming the presence of a background radio continuum emitter (on-source). The blue line is the observed absorption fraction $e^{-\tau_{obs}}$, while the red dash lines show a decomposition of the absorption spectrum into Gaussian components.

by iteratively evaluating

$$I_{\text{HI},v,i+1} = e^{-\kappa_{v,i}\,\Delta L} I_{\text{HI},v,i} + j_{v,i}\Delta L, \qquad (12)$$

for each velocity channel, where $i$ again runs for 1 to 200. Our final emission spectrum is the intensity $I_{\text{HI}}$ that emerges from the last pixel.

In Fig. 4, we show an example result of applying this procedure to the LOS shown in Fig. 3; in the figure we plot the brightness temperature $T_B = \lambda_a^2 I_{\text{HI}}/(2k_B)$ (top) and absorption fraction $e^{-\tau}$ (bottom) as functions of the channel relative velocity $\Delta V_{LOS} \equiv V_c - \bar{V}_{LOS}$.

## 3.3 Emission and B-field maps

After solving equation (12) for all LOSs, we can integrate the derived emission spectra over velocity to produce two maps of total H I 21 cm line intensity, $I_{\text{HI}}$, normal to $x$- and $y$-axis, respectively. We similarly produce maps of OH velocity-integrated intensity, $I_{\text{OH}}$, averaging the intensities of the 1665 and 1667 MHz lines. We compare the resulting H I and OH total intensity maps to the true column density map for cloud 4 (as shown in Fig. 2) in Fig. 5; the true column density $N_{\text{H}}$ we show is integrated over the same 100-pc lines of sight used for the simulated emission maps. The three maps do not perfectly mimic each other. The $N_{\text{H}}$ and $I_{\text{HI}}$ maps differ because low $N_{\text{H}}$ regions can be ionized and high $N_{\text{H}}$ regions can be optically thick and can be dominated by $H_2$, all of which decrease the observed intensity. The OH map strongly picks out only the densest region, and due to varying excitation the OH intensity distribution does not perfectly

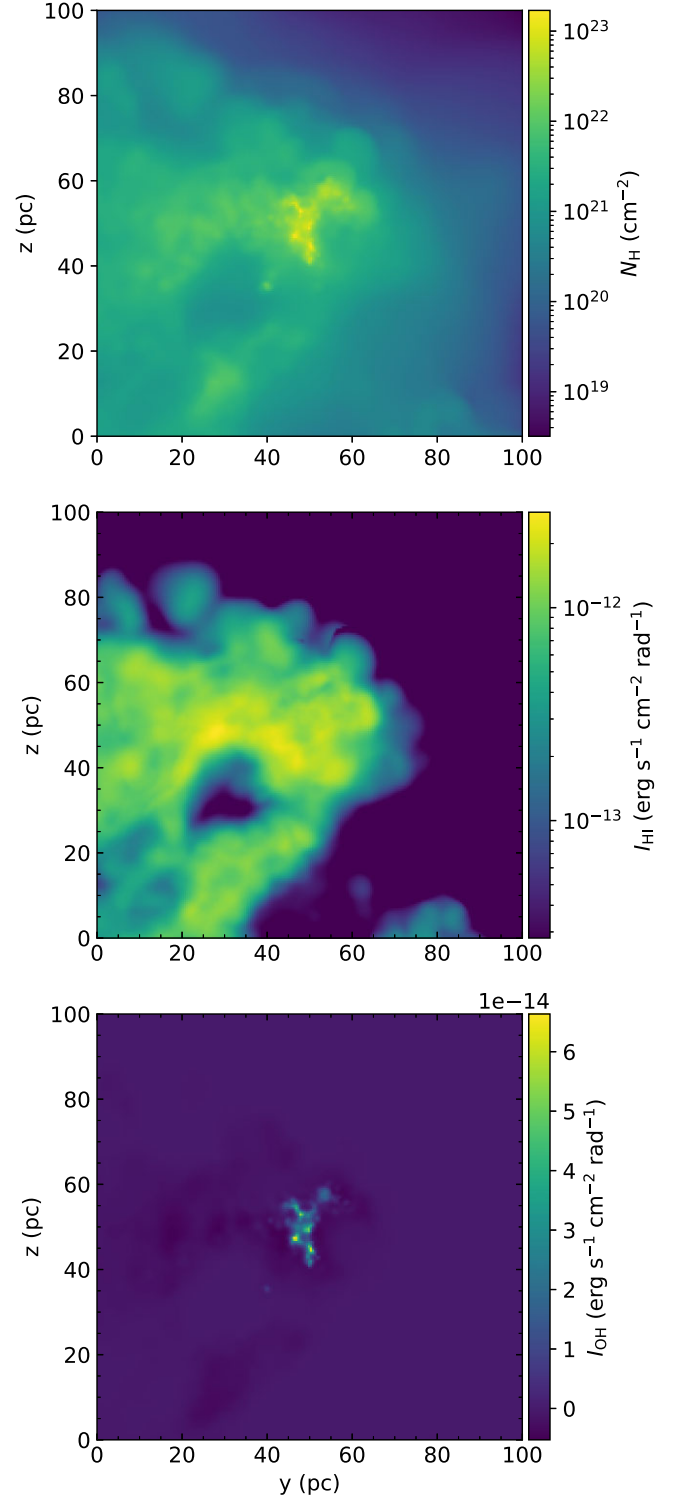

**Figure 5.** Maps from the $x$-axis projection of the $(100 \text{ pc})^3$ cube extracted around the densest gas particle in cloud 4. Top panel: total column density of hydrogen nuclei $N_{\text{H}}$. Middle panel: H I emission line velocity-integrated intensity $I_{\text{HI}}$. Bottom panel: total intensity $I_{\text{OH}}$ in the OH 1665 and 1667 MHz lines; the quantity we plot is the average of the two lines.

mirror the total density distribution even in regions where there is strong OH emission.[2]

Our final task is to derive magnetic field measurements from the synthetic spectra. We do so using three of the Zeeman techniques described in Section 1: OH emission, H I absorption against a background radio source, and H I narrow self-absorption (HINSA).

When one uses OH emission lines to measure the magnetic field strength through the Zeeman effect, the emission regions that contribute more photons to the observer will contribute more to the measured $B$-field strength. Since the lines are optically thin and single-peaked, the contribution is simply linear in the emission. Therefore, we approximate the magnetic field that would be inferred from the OH lines as simply the $L_{OH}$-weighted mean LOS magnetic field for both OH 18 cm lines. We then take the average of these two values as the final result $B_{OH}$. In Fig. 6, we compare the true, volume-weighed mean LOS magnetic field for cloud 4 projected along the $x$-axis with maps generated from synthetic OH observations. For $B_{OH}$ we show only pixels where $I_{OH} > I_{OH, max}/100$ as a rough proxy for where the OH lines would be bright enough to be detectable.

We next consider H I absorption against a background radio continuum source, which is somewhat more complex since each H I absorption spectrum typically contains several cold neutral medium components with different $V_{LOS}$. To make separate measurements for each component, following the approach of Heiles & Troland (2004), we perform least-squares fits of at most five Gaussians to the absorption spectrum. The optical depth of a single-component $\tau_{comp, \nu}$ is

$$\tau_{comp, \nu} = \frac{\tau_{norm}}{\sqrt{2\pi\sigma_v^2}} e^{-(V_c - V_{comp})^2/(2\sigma_v^2)}, \tag{13}$$

where $\tau_{norm}$, $\sigma_v$ and $V_{comp}$ are parameters to be fit. We show the example of such a multicomponent Gaussians as red dash lines in the bottom panel of Fig. 4. Due to computational resource limitations, we do not perform this decomposition for all of our $5.6 \times 10^5$ spectra; instead, for each projection map we decompose 1000 randomly selected LOSs, chosen out of the set with $N_{HI} \in (10^{19}, 10^{21.5})$ cm$^{-2}$, the same column density range observed in Heiles & Troland (2004).

Once we have decomposed the H I spectra, we next compute the synthetic Zeeman-inferred magnetic field for each. This requires some subtlety, because components in velocity space do not necessarily map on to real structures in physical space, as is implicitly assumed by the decomposition procedure used by Heiles & Troland (2004) and others. In such situations, the marginal contribution of the magnetic field in a particular fluid element in space to the magnetic field inferred for a particular Gaussian component will depend to some extent on the details of the decomposition and fitting procedure. However, generically we expect that the contribution of a particular voxel to the magnetic field inferred for a particular component to scale with both the total amount of absorption it contributes and with its marginal contribution to that particular component. We there define a weight $w_i$ for each voxel $i$'s contribution to a particular

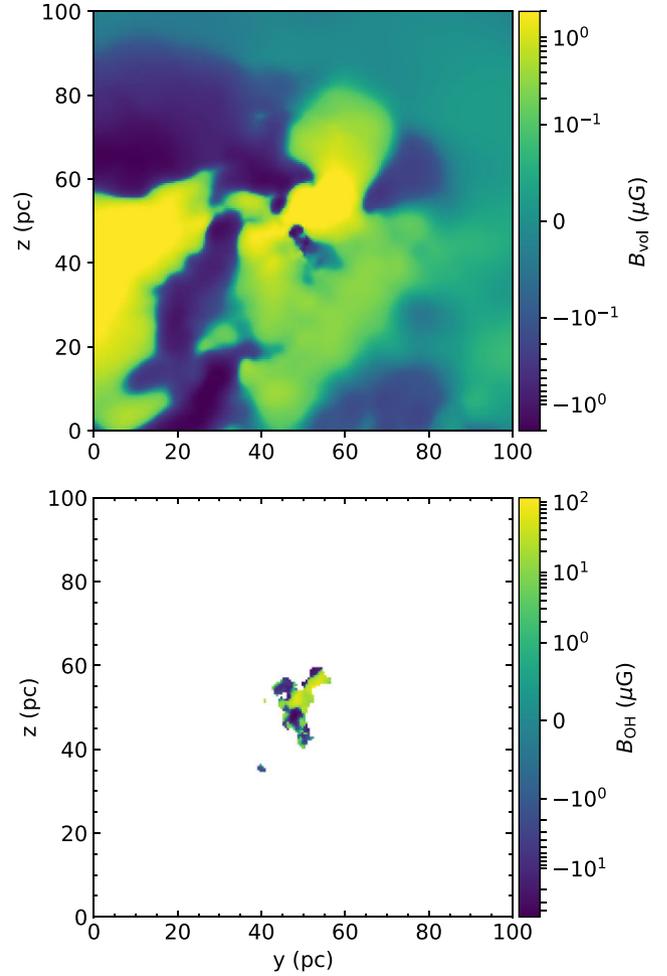

**Figure 6.** Line-of-sight magnetic field maps from the $x$-axis projection of the $(100\,pc)^3$ cube extracted around the densest gas particle in cloud 4. Top panel: true, volume-weighted mean line-of-sight magnetic field $B_{vol}$ map. Bottom panel: magnetic field inferred from synthetic OH Zeeman observations, $B_{OH}$. Note that we only show $B_{OH}$ for pixels with total OH intensity $I_{OH} > I_{OH, max}/100$. Also note that the colourbar range differs in the two panels.

component of central velocity $V_{comp}$ and dispersion $\sigma_v$ as

$$w_i = e^{-(V_{LOS,i} - V_{comp})^2/(2\sigma_v^2)} \int k_{\nu,i} d\nu, \tag{14}$$

and the resulting synthetic $B$-field measurement for one Gaussian component as

$$B_{HI,comp} = \frac{\sum_i w_i B_{LOS,i}}{\sum_i w_i}. \tag{15}$$

We only perform the calculations above for Gaussian components with $\sigma_v < 2$ km s$^{-1}$, to make sure that most of the component lies within the spectral range we have extracted. From all selected LOSs we find 30 853 components and derive $B_{HI,comp}$ values for them.

Our final task is to make synthetic HINSA magnetic field measurements. Since HINSA only occurs where there is cold H I ($T \lesssim 100$ K), we first project only cold H I into column density maps, then find the pixel with the highest column density, and determine the $B$-field in the cold H I $B_{HI,cold}$ from the 100-pc long LOS projected on to this pixel. From Fig. 3 we can see that the cold H I region has nearly constant $V_{LOS}$, thus we can assume the Doppler profile shape

---

[2] Careful readers may note that the OH intensity $I_{OH}$ shown in the bottom panel of Fig. 5 can be slightly negative. This is a real effect, which occurs because differential selection rules for the upper and lower hyperfine states of the OH 1665 and 1667 MHz lines can drive the excitation temperatures below the CMB temperature, so the line is seen in absorption against the CMB. As a result the total intensity of the line, after subtracting off the CMB, can be slightly negative along some lines of sight.

is constant for the whole cold H I region, and determine $B_{\rm HI, \, cold}$ as

$$B_{\rm HI,cold} = \frac{\sum_{\rm i,cold} B_{\rm LOS,i} \int k_{\nu,i} \, d\nu}{\sum_{\rm i,cold} \int k_{\nu,i} \, d\nu}, \qquad (16)$$

where $\sum_{\rm i,cold}$ means summation over pixels with $T \leq 100$ K. After two projections per cube, we have 14 $B_{\rm HI, \, cold}$ results. Note that we make only one measurement towards each cloud in order to mimic the approach of the published HINSA measurement (Ching et al. 2022), which is typically conducted along the LOS through the HINSA maximum.

It is clear that the B-field strength and morphology highlighted by OH concretely differ from the mean B-field over the 100-pc-long LOS, indicating a possible inconsistency between the observed mass-to-flux ratio and the real overall magnetic-gravity energy relation. Since for each LOS we make several $B_{\rm HI}$ measurements from the fitted Gaussian components, individually, and the selected LOSs do not cover the whole $(100 \, {\rm pc})^2$ area, we cannot show a $B_{\rm HI}$ map similar to Fig. 6.

## 4 RESULTS

In this section, we first reproduce the observed *B–N* relation with synthetic observations in Section 4.1, then explore how well this relation captures the true ratio of gravitational to magnetic energy from the full 3D simulation data in Section 4.2.

### 4.1 Magnetic field versus column density

After making the synthetic observations, we can now compare magnetic field to column density following Crutcher (2012), and look for the magnetic criticality transition. The pixel size is 0.5 pc in both Fig. 5 and Fig. 6, which is close to the 3 arcmin beam width of the Arecibo telescope when applied to a cloud at a distance of $\sim$300 pc. Therefore, we simply treat each pixel in the OH emission map as a synthetic beam, and directly take OH-inferred magnetic field values $B_{\rm OH}$ from the pixel values derived in Section 3.1. We similarly derive the corresponding column density from the emission intensity maps. Assuming the gas is optically thin (as is assumed in observations), we convert OH intensity into hydrogen column density as

$$N_{\rm H,OH} = \frac{a \lambda^3 I_{\rm OH}}{2 k_{\rm B}}, \qquad (17)$$

where $\lambda$ is the emission line wavelength, $I_{\rm OH}$ is the velocity-integrated intensity, and $a$ is an empirical constant. For the two OH 18 cm lines, we have $a = 2.12 \times 10^{22}$ cm$^{-2}$/(K km s$^{-1}$) for the 1665 MHz line and $a = 1.18 \times 10^{22}$ cm$^{-2}$/(K km s$^{-1}$) for the 1667 MHz line, respectively (Troland & Crutcher 2008).

We again follow observational procedure in deriving the column density related to the H I absorption in each Gaussian component. For an isothermal cloud, the H I column density is related to the velocity-integrated optical depth by (e.g. Draine 2011)

$$N_{\rm HI} = \left( 1.82 \times 10^{18} \frac{\rm cm^{-2}}{\rm K \, km \, s^{-1}} \right) T_{\rm s,comp} \int \tau_{\rm comp,\nu} \, dV, \qquad (18)$$

where $T_{\rm s, \, comp}$ is the spin temperature, and the integral is over the velocity range of the absorption spectrum. With the combination of the off-source emission spectrum $T_{\rm B}$ and the on-source absorption spectrum $\tau$, we can determine $T_{\rm s, \, comp}$ as

$$T_{\rm s,comp} = \frac{T_{\rm B,peak}}{1 - \exp(-\tau_{\nu,peak})}, \qquad (19)$$

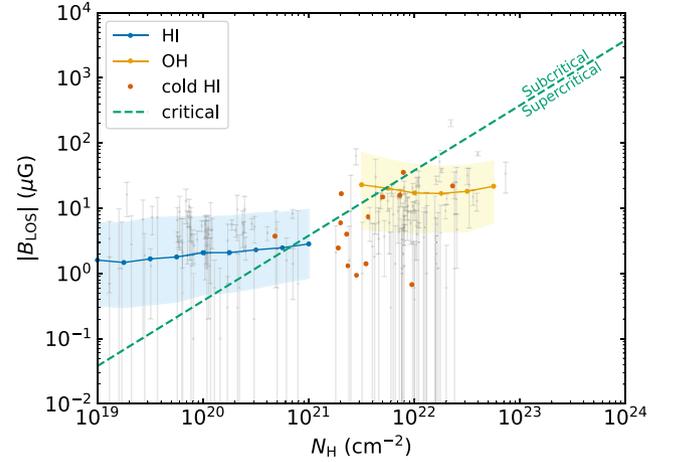

**Figure 7.** Magnetic field strengths inferred from synthetic Zeeman effect observations using H I absorption against background radio sources (blue band), OH emission (yellow band), and H I narrow self-absorption (red points), as a function of inferred column density. For the yellow and blue bands, lines with points show 50th percentile values of the $|B_{\rm LOS}|$ in bins of column density 0.25 dex wide, while the band edges illustrate the 16th to 84th percentile range. The grey points show real observations, from the compilation of Crutcher (2012). The green dashed line represents $\mu_B = 1$ (equation 20). Measurements below this line correspond to supercritical conditions, while those above are sub-critical.

where the subscript *peak* indicates values measured at the centre of the Gaussian component where the line-of-sight velocity $V_{\rm LOS} = V_{\rm comp}$. This gives column densities for the H I absorption against background radio source measurements.

Finally, since we did not produce synthetic spectra for the HINSA observations, we simply take the hydrogen column density in the form of $H_2$ as $N_{\rm H, \, cold}$, the approach matches that used in Ching et al. (2022), though in their case they derive the $H_2$ column from an independent estimate. In our case we take the $H_2$ column directly from the simulation data.

For consistent comparison with Crutcher (2012), we select H I synthetic beams with a converted column density $N_{\rm HI} \in (10^{19}, \; 10^{21})$ cm$^{-2}$, and OH synthetic beams with a converted column density $N_{\rm H,OH} > 10^{21.5}$ cm$^{-2}$. In Fig. 7, we show the median and 16th to 84th percentile range of measured magnetic field strength from synthetic H I absorption and OH measurements in bins of column density 0.25 dex wide; the blue band shows the former, the yellow band shows the latter. We also plot the 14 HINSA measurements towards the densest point in each synthetic map as individual red points. The dashed green line shows the sub-/supercritical transition, defined by

$$|B_{\rm LOS,crit}| = 3\pi \sqrt{\frac{G}{5}} \, \mu m_{\rm H} N_{\rm H}, \qquad (20)$$

where $\mu = 1.4$ is the mean mass per H nucleon for gas with the usual composition of $\approx 75$ per cent H and $\approx 25$ per cent He by mass. Finally, we show the observations compiled by Crutcher (2012) as grey dots with error bars for comparison.

The primary point to take from Fig. 7 is that our synthetic observations do an excellent job of reproducing the real observations. The median LOS field strength is $\sim$1 μG for H I observations, and rises to $\sim$10 μG for OH observations. We can also see a clear transition in magnetic criticality at column densities $N_{\rm H} \in (10^{21}, \; 10^{22})$ cm$^{-2}$, such that the majority of the H I absorption measurements indicate sub-criticality, while the OH measurements are almost exclusively in the supercritical region. Again consistent with the result of Ching et al. (2022), most HINSA synthetic observations also lie in the supercritical region.

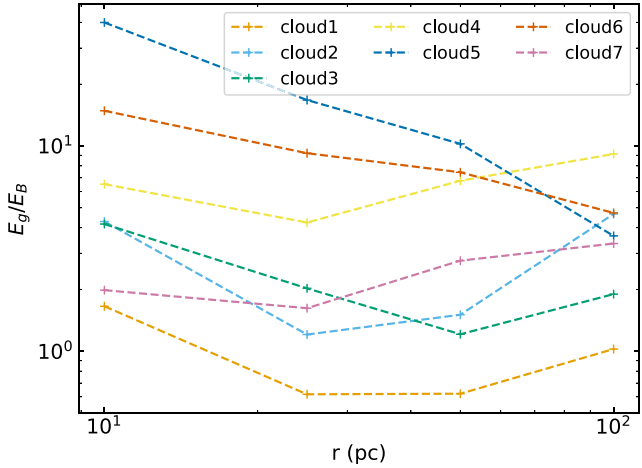

**Figure 8.** Ratios of gravitational energy $E_g$ to magnetic energy $E_B$ measured in spheres with different radii $r$, centred on the densest point in each sample cloud.

### 4.2 Gravitational versus magnetic energy

Although the synthetic Zeeman effect measurements agree with previous observational works, these may not reveal the true relation between gravitational energy $E_g$ and magnetic energy $E_B$. To investigate these, we compute the total gravitational potential and magnetic energies within some volume of interest $V$ as

$$E_g = -\frac{1}{2} \sum_p \sum_{q,\, q \neq p} \frac{G M_p M_q}{r_{pq}}, \tag{21}$$

$$E_B = \int_V \frac{B^2}{8\pi} \, dV. \tag{22}$$

In the equations above, $G$ is the gravitational constant, $p$ and $q$ are indices that run over all the GIZMO particles within $V$, $M_p$ and $M_q$ are the particle masses, $r_{pq}$ is the distance between particle $p$ and particle $q$, and $B$ is the local total magnetic field strength. We evaluate these energies as a function of size scale by identifying the densest gas particle in each of our seven sample clouds, then drawing four spheres with radii $r = 10, 25, 50,$ and $100$ pc. We choose these size scales to span the atomic-to-molecular transition that is, based on the observational diagnostics, closely related to the sub-/supercritical transition: for all seven samples, the $H_2$ mass fraction is $\approx 0.9$ for $r < 10$ pc, but drops to around 0.3 for 90 pc $< r < 100$ pc, indicating that our sample regions span from the molecular core to the cloud edge. The 50 pc sphere is also close in size to the 100 pc region on to which we interpolate on to a grid when computing the synthetic Zeeman effect, though we will see below that the synthetic observations are actually sensitive to considerably smaller scales than this.

In Fig. 8, we plot the $E_g/E_B$ versus sphere radius for all seven clouds. It is clear that $E_g/E_B > 1$ for all clouds on most scales. The only exception is cloud 1, which has $E_g/E_B \approx 0.6$ on $r = 25$ and 50 pc scales, which may be related to the fact that it has the smallest total mass in our sample; however, even these sub-critical regions are embedded within a larger 100 pc-scale supercritical one. Therefore, from the 3D information, we conclude that all seven clouds are already supercritical on scales where the H I synthetic observations seem to indicate that they are sub-critical. Using the full

3D information, we find no transition from a sub- to a supercritical state across the H I – $H_2$ transition. This suggests that the observed transition in magnetic criticality across the column density is most likely a result of inconsistency between 2D projected observations and real 3D structures. We explore the nature of this mismatch in the next section.

## 5 DISCUSSION

We have seen that, although both actual Zeeman measurements and our synthetic versions of them indicate the existence of a transition between sub- and supercritical states as material transitions from H I to $H_2$, no such transition is actually found in our simulations; instead, even on 100 pc scales around giant molecular clouds where the mass is predominantly atomic, clouds are supercritical. Our goal in this discussion is to understand why the H I Zeeman observations seem to indicate the presence of a sub-critical state where none actually exists. We find that this sub-critical illusion has two contributing factors. First, in previous observational work, the column density $N$ associated with a particular H I Zeeman measurement is directly converted from the observed absorption line depth, which measures only H I material, and misses other chemical states of gas. Second, Zeeman effect measurements only provide information on the magnetic field in regions of strong absorption (or emission for OH), and such regions may not be representative of the true magnetic field strength, leading to a biased estimate of $E_g/E_B$.

Starting with the first of these factors, to study how accurate the $N_{\rm HI}$ synthetic measurement is, we compare it with results directly derived from the simulation data. For each Gaussian component for which we make a synthetic Zeeman measurement, we first find the related LOS, then compute the density $n_{\rm HI}$ at positions along that LOS where the LOS gas velocity $V_{\rm LOS}$ lies in the range $V_{\rm comp} \pm \sigma_{v,\rm comp}$, i.e. we find all the fluid elements whose velocities lie within one standard deviation of the central velocity of each component. We then compute the column density $N_{\rm HI,1\sigma_v}$ of these fluid elements. We then define the intrinsic H I column density related to the Gaussian component as $N_{\rm HI,sim} = N_{\rm HI,1\sigma_v}/0.683$, where 0.683 is the fraction of the mass that lies within $\pm 1\sigma$ of the centre for a Gaussian distribution. We also use an identical procedure to derive the total hydrogen column density $N_{\rm H,sim}$ by projecting the total density of H nuclei $n_{\rm H}$; $N_{\rm H,sim}$ differs from $N_{\rm HI,sim}$ in that the former includes all hydrogen regardless of its chemical state, while the latter includes only hydrogen that is chemically in the form of H I.

We can compare these true column densities to the columns $N_{\rm HI}$ derived from the 21 cm spectra using observational methods, as described in Section 4.1, equations (15) and (19). The ratios $(N_{\rm HI}/N_{\rm HI,sim})$ and $(N_{\rm HI}/N_{\rm H,sim})$ then tell us how far the observational measurements lie from the H I mass and the total mass. In Fig. 9, we plot the 16th, 50th, and 84th percentiles of $(N_{\rm HI}/N_{\rm HI,sim})$ and $(N_{\rm HI}/N_{\rm H,sim})$, binning the data by observationally estimated column density $N_{\rm HI}$ in the same pattern as in Fig. 7. From the figure we can see that the observational procedure recovers about $\sim 40$ per cent of H I mass, but only $\sim 10$ per cent of the total mass. The reason is that most mass along the LOS is in the form of $H_2$ and ionized hydrogen. In retrospect this result is not surprising: observational probes of H I around molecular clouds do not represent a randomly selected sample of lines of sight through the ISM. Instead, they are probing regions where it is reasonable to believe that dark gas composed mostly of $H_2$ might exist. This result shows that observationally inferred mass-to-flux ratios likely suffer from underestimates of $N_{\rm HI}$ due to chemical composition variations.

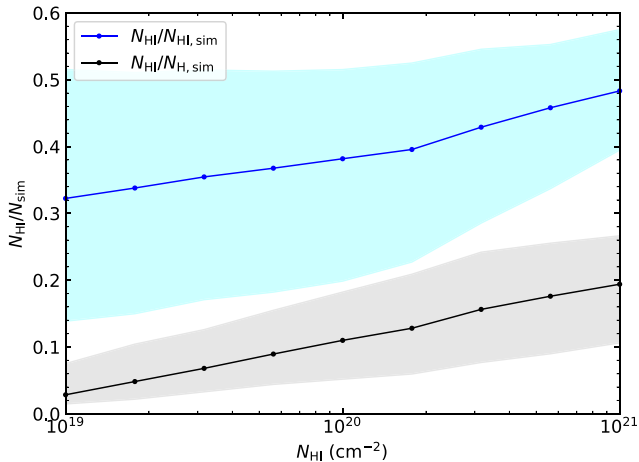

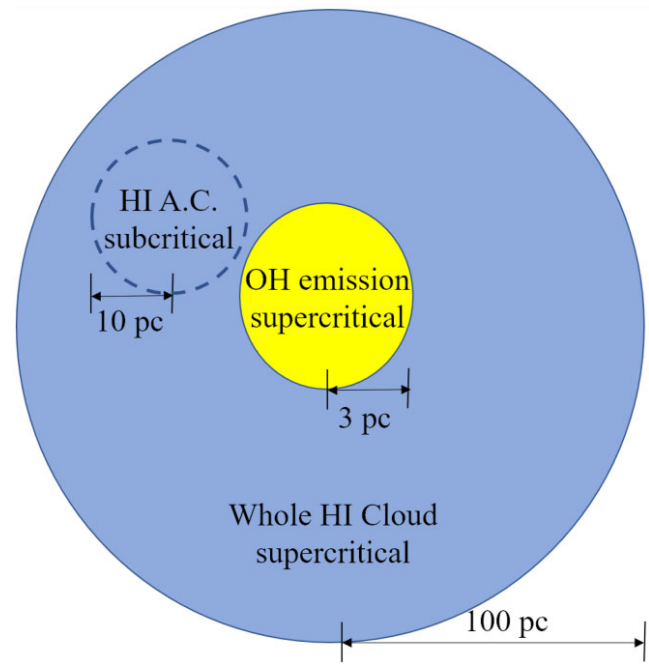

**Figure 9.** The blue band shows the ratio of inferred to true H I column density, ($N_{HI}/N_{HI, sim}$), while the grey band shows the ratio of inferred H I column density to true column density of H nuclei in all chemical states, ($N_{HI}/N_{H, sim}$). For each band, lines with dots show the 50th percentile values in bins of observationally inferred H I column density $N_{HI}$, while the band edges illustrate the 16th and 84th percentiles.

The second factor contributing to the inconsistency between the observed and true levels of magnetic criticality we identify in Section 4 is the limited spatial scale of the magnetic field probed by Zeeman effect measurements. From equation (15) we learn that $B_{HI}$ measurements are dominated by regions with large absorption coefficients, which may only reveal only a small part of the overall magnetic field structure. To quantify this effect, we compute an effective length-scale probed by H I absorption measurements as follows. For each LOS, we assign the $i$th sample point on that LOS a position $x_i = i\Delta L$. We then define the length-scale of H I the B-field measurement $l_{HI}$ to be the weighted standard deviation of $x_i$, where the weight $w_i$ is defined by that sample point's contribution to the synthetic Zeeman measurement, as defined in equation (14). Quantitatively, we compute $l_{HI}$ as

$$x_{mean} = \frac{\sum_i w_i x_i}{\sum_i w_i}, \quad (23)$$

$$l_{HI} = \left[ \frac{\sum_i w_i (x_i - x_{mean})^2}{\sum_i w_i} \right]^{1/2}, \quad (24)$$

where the sum runs over all the sample points $i$ along a given LOS. Intuitively, $l_{HI}$ calculated in this way quantifies the length along the LOS of the region that contributes to a Zeeman measurement; for uniform $w_i$, given our 100 pc sampling region, $l_{HI}$ equals 57.7 pc, while the smallest $l_{HI}$ that we can resolve given our sampling, corresponding to a case where the region that produces the absorption is captured by exactly two sample points that both contribute equally, is 0.25 pc. Performing this calculation over all LOS's, we find that the 16th, 50th, and 84th percentiles of $l_{HI}$ are 4.11 pc, 11.4 pc, and 29.3 pc, respectively. Therefore, our synthetic H I observations generally measure magnetic fields in regions of strong H I absorption whose characteristic size scale is ∼10 pc. We have performed a similar calculation on OH emission as in equation (24), but we change the weight function to $L_{OH}$, and derive the characteristic size scale of OH Zeeman effect measurement as ∼3 pc.

To see why selecting H I-dominated regions of this size scale might produce a bias, for each (100 pc)$^3$ cube we identify all sample points where at least 75 per cent of hydrogen atoms are in the form of H I; then we randomly select 20 such points that are at least 10 pc from each other, and draw $r = 10$ pc spheres around them. We determine the $E_g/E_B$ ratio for these spheres, yielding a total of 140 samples – 20 from each of the seven sample clouds. For these samples we find a median $E_g/E_B = 0.25$, and the ratio is less than unity for 63 per cent of them. This result indicates that ∼10-pc scale regions of H I-dominated gas are likely to be locally sub-critical, which is consistent with the observations of Heiles & Troland (2004) and others. However, this is an illusion created by sampling unrepresentative regions: when considering the whole cloud from which these H I regions are drawn on 100 pc scales, or when selecting regions centred on the density maximum rather than maxima of H I absorption, we find that the gas is supercritical, as shown in Fig. 8.

**Figure 10.** Schematic view of regions detected by Zeeman effect observation in the H I cloud. The dashed circle labelled as H I A.C. is the H I sub-region measured by an absorption component (A.C.) in observation. Note that the length-scale of different regions is not strictly proportional.

This analysis leads us to propose a scenario for the relation between gravity and magnetic fields in molecular clouds and their neutral hydrogen surroundings illustrated in Fig. 10. For a cloud ∼100 pc in size, its dense molecular core is supercritical, and is revealed as such when it is measured using OH emission lines. Small H I sub-regions in the cloud outskirts are generally locally sub-critical, in the sense that their magnetic energy is larger than their gravitational *self*-energy, consistent with the H I absorption observations. However, even for regions that are genuinely sub-critical on ∼10 pc scales, if we consider the large-scale cloud structure as a whole, the molecular mass ignored in H I observations can greatly increase the total gravitational energy and make the whole cloud supercritical. Magnetic fields may be able to locally support H I regions against gravity, but cannot resist gravitational collapse on the large scale. One can make a useful analogy here to the competition between thermal and gravitational energy in the envelope of the Sun – if one considers only the outer layers of the Sun, say the outer half in radius, its thermal energy is far greater than its gravitational binding energy. However, this is a misleading statistic, because the dominant source of gravitational binding for the outer 20 per cent of the Sun is not its self-energy, but instead the gravitational attraction provided by the remaining 80 per cent of the mass. A measurement that is *only* sensitive to the ratio of thermal to gravitational energy in the envelope will completely miss this effect.

In this view, the apparent transition between sub-critical and supercritical gas seen in Zeeman measurements, and described in Crutcher (2012), is a confusion caused by mixing local and cloud–scale energy relations. Correct recovery of the level of magnetic criticality requires the ability to measure both the magnetic field and the true column across a whole molecular cloud. Zeeman measurements do not provide this ability.

# 6 CONCLUSIONS

This work aims at understanding the apparent transition in magnetic criticality between H I- and H$_2$-dominated regions found in Zeeman effect observations. Such study requires a combination of high-resolution simulations of the molecular cloud formation process and synthetic observations that capture the complexities of Zeeman measurements in different chemical tracers, following real observational procedures as closely as possible. To this end, we perform the first MHD simulation of molecular cloud formation out of realistic full-galactic magnetic field structure. Starting from the low-resolution galactic simulation, we zoom into seven molecular clouds, then trace and re-simulate their formation history at high resolution (∼0.4 pc). We then post-process the simulations to derive realistic chemical compositions, gas temperatures, and non-LTE level populations, yielding physically accurate calculations of emissivities and absorption coefficients in the most commonly used Zeeman-sensitive lines. We use these to solve the radiative transfer equation along 100-parsec-long lines of sight to produce synthetic H I 21 cm line absorption and emission spectra, which mimic the absorption/emission strategy used in real H I Zeeman measurements, as well as synthetic OH emission and H I narrow self-absorption (HINSA) observations. We process these results using the same procedures used in observations to produce realistic synthetic measurements of magnetic field strengths and column densities in different tracers.

We find that our synthetic *B* versus *N* relations are quantitatively consistent with the observed trends summarized by Crutcher (2012): molecular clouds appear to be sub-critical in H I observations, and become supercritical in OH observations; the transition of criticality happens in the hydrogen column density range of $N_H \in [10^{21}, 10^{22}]$ cm$^{-2}$. HINSA measurements probe this column density regime, and generally indicate that clouds are supercritical.

Such results, however, may not reveal the real ratio between gravitational energy and magnetic energy $E_g/E_B$. We compare out synthetic Zeeman measures to ratios of $E_g/E_B$ computed directly from the 3D simulation data, evaluated over a range of size scales and for a sample of different molecular clouds. On scales from 10 to 100 pc we find that the real ratio $E_g/E_B > 1$ for most clouds, which indicates that gravity dominates over magnetic fields. This applies even when averaging over size scales large enough that the material we are probing is largely atomic, indicating that the H I Zeeman observations do not accurately reflect the true dynamical state of the gas.

We identify two main contributors to this inconsistency. First, column densities used in H I Zeeman observations only measures the hydrogen mass in the neutral atomic form, which underestimates the true total column density along lines of sight where the chemical composition is not uniformly atomic hydrogen. Second, the regions probed by H I absorption Zeeman measurements capture density and magnetic field structure in ∼10 pc-scale regions on the outskirts of molecular clouds. Although these local H I sub-regions may be sub-critical as observed, they are none the less part of larger structures that are still supercritical once we include the gravitational influence of the central dense molecular region. We conclude that accurate observational determination of the relative importance of magnetic fields and gravity in the molecular cloud formation process requires careful attention to ensure that the observations accurately capture all the self-gravitating mass, and that the regions captured by observations are representative of entire clouds, rather than small sub-regions within them.

## ACKNOWLEDGEMENTS

The authors acknowledge Prof. Di Li, Dr Tao-chung Ching, and Dr Hiep Nguyen for suggestions on synthetic observations. We further acknowledge funding from the Australian Research Council through its Future Fellowship and Laureate Fellowship funding schemes, awards FT180100375 and FL220100020, and high-performance computing resources provided by the Australian National Computational Infrastructure (grants jh2 and ek9) through the National and ANU Computational Merit Allocation Schemes, and by the Leibniz Rechenzentrum and the Gauss Centre for Supercomputing (grant pr32lo).

## DATA AVAILABILITY

The data underlying this article will be shared upon reasonable request to the corresponding author.

This paper has been typeset from a TeX/LaTeX file prepared by the author.